\documentclass[aps,pra,onecolumn,superscriptaddress,notitlepage]{revtex4-1}

\usepackage{xcolor,graphicx}
\usepackage{amssymb,amsmath,graphicx}
\usepackage[utf8]{inputenc}

\usepackage{caption}
\usepackage{subcaption}
\usepackage{sidecap}
\usepackage{wrapfig}

\newcommand{\mus}{\mu{\rm s}}
\newcommand{\kHz}{{\rm kHz}}
\newcommand{\MHz}{{\rm MHz}}

\usepackage{color}

\begin{document}

\title{Experimental realization of optimal time-reversal on an atom chip for quantum undo operations}

\author{Ivana Mastroserio}
\affiliation{\mbox{Dipartimento di Fisica e Astronomia, Universit\`{a} di Firenze,} via Sansone 1, I-50019 Sesto Fiorentino, Italy.}
\affiliation{\mbox{LENS, Universit\`{a} di Firenze,} via Nello Carrara 1, I-50019 Sesto Fiorentino, Italy.}
\affiliation{Dipartimento di Fisica ``Ettore Pancini'', Universit\`{a} degli Studi di Napoli Federico II, Napoli, Italy.}

\author{Stefano Gherardini}
\affiliation{\mbox{LENS, Universit\`{a} di Firenze,} via Nello Carrara 1, I-50019 Sesto Fiorentino, Italy.}
\affiliation{Istituto Nazionale di Ottica (CNR-INO), Area Science Park, Basovizza, I-34149 Trieste, Italy}

\author{Cosimo Lovecchio}
\affiliation{\mbox{Dipartimento di Fisica e Astronomia, Universit\`{a} di Firenze,} via Sansone 1, I-50019 Sesto Fiorentino, Italy.}
\affiliation{\mbox{LENS, Universit\`{a} di Firenze,} via Nello Carrara 1, I-50019 Sesto Fiorentino, Italy.}

\author{Tommaso Calarco}
\affiliation{\mbox{Peter Gr\"{u}nberg Institute -- Quantum Control (PGI-8)}, Forschungszentrum J\"{u}lich, J\"{u}lich, Germany.}

\author{Simone Montangero}
\affiliation{\mbox{Dipartimento di Fisica e Astronomia ``G. Galilei'' \& Padua Quantum Technologies Research Center, Universit\`{a} di Padova}, \\ I-35131 Italy.}
\affiliation{\mbox{Istituto Nazionale di Fisica Nucleare (INFN)}, Sezione di Padova, I-35131 Padova, Italy.}

\author{Francesco S. Cataliotti}
\affiliation{\mbox{Dipartimento di Fisica e Astronomia, Universit\`{a} di Firenze,} via Sansone 1, I-50019 Sesto Fiorentino, Italy.}
\affiliation{\mbox{LENS, Universit\`{a} di Firenze,} via Nello Carrara 1, I-50019 Sesto Fiorentino, Italy.}
\affiliation{Istituto Nazionale di Ottica (CNR-INO), Largo Enrico Fermi 6, I-50125 Florence, Italy.}

\author{Filippo Caruso}
\affiliation{\mbox{Dipartimento di Fisica e Astronomia, Universit\`{a} di Firenze,} via Sansone 1, I-50019 Sesto Fiorentino, Italy.}
\affiliation{\mbox{LENS, Universit\`{a} di Firenze,} via Nello Carrara 1, I-50019 Sesto Fiorentino, Italy.}
\affiliation{Istituto Nazionale di Ottica (CNR-INO), Largo Enrico Fermi 6, I-50125 Florence, Italy.}

\begin{abstract}
We report on the use of the dCRAB optimal control algorithm to realize time-reversal procedures for the implementation of quantum \emph{undo} operations, to be applied in quantum technology contexts ranging from quantum computing to quantum communications. By means of the \emph{undo} command, indeed, the last performed operation can be time-reversed so as to perfectly restore a condition in which an arbitrary new operation, chosen by the external user, can be applied. Moreover, by further generalizing this concept, the \emph{undo} command can also allow for the reversing of a quantum operation in a generic instant of the past. Here, thanks to optimal time-reversal routines, all these functionalities are experimentally implemented on the five-fold $F=2$ Hilbert space of a Bose-Einstein condensate (BEC) of non-interacting $^{87}$Rb atoms in the ground state, realized with an atom chip. Specifically, each time-reversal transformation is attained by designing an optimal modulated radio frequency field, achieving on average an accuracy of around $92\%$ in any performed test. The experimental results are accompanied by a thermodynamic interpretation based on the Loschmidt echo. Our findings are expected to promote the implementation of time-reversal operations in a real scenario of gate-based quantum computing with a more complex structure than the five-level system here considered.
\end{abstract}

\maketitle

\section{Introduction}

The \emph{undo} command allows to reverse an operation that has been performed in a past step of a complex computational routine. Specifically, the \emph{undo} command is a basic tool to be addressed in all those computational processes, in which the external user may need to proceed step by step, thus visualizing the result of each operation. This already holds in classical computer or computing systems managed by a high-level interface (as e.g.\,an operating system) where such a command is a requirement that is practically taken for granted \cite{BerlageTCHI1994,JakubecPCS2014}.

In quantum platforms for quantum computing, the \emph{undo} command is expected to need more onerous procedures with respect to the classical case. So far, procedures carrying out time-reversal transformations \cite{Oreshkov} have been implemented in superconducting circuits realizing quantum circuits \cite{Lesovik}, and classical \cite{Cohen} and quantum optics platform \cite{SchianskyArXiv2022}. As a main feature, it is desirable to employ universal features that are valid in a general quantum technologies context. In this regard, two main challenges have to be still addressed: one from the procedural/numerical side, ensuring high performance and high speed, and the other from a technological/experimental point of view. In our view, these challenges concerns the establishment of an optimal procedure for the realization of quantum \emph{undo} operations by reducing as much as possible the execution error and requiring a moderate computation load depending on the experimental devices at disposal.

In real experiments, indeed, one can often implement only a small set of operations, due to practical limitations, experimental imperfections and restrictions on resources. Moreover, such an optimal procedure has to be designed to be possibly implemented in a generic experimental platform. The optimal solution to these issues, which we are going to propose, is the use of quantum Optimal Control (OC) methods that have been introduced for the control of quantum systems dynamics. Quantum OC theory is one of the optimal ways to successfully prepare quantum states and perform desired tasks that are crucial in implementing quantum-based technologies, ranging from atomic, molecular and optical systems to solid-state systems \cite{SB1997,NelsonPRL2000,chu2002,WR2003,SteckPRL2004,Mabuchi,DAlessandroBook2007,Rabitz,WisemanBook,BrifNJP2010,AltafiniTAC2012,C2014,Koch2016,RossiNature2018,GherardiniBattery2019,GirolamiPRL2019,Mueller2020,Gherardini_QQcontrol}. Specifically, in this paper we adopt the dressed Chopped Random Basis (dCRAB) optimal control algorithm \cite{CRAB,DCM,CanevaPRA2014,RachPRA2015,MuellerReview2021} that has been already successfully tested in several experiments even involving many-body atomic systems \cite{R2013,Frank14,nostroopt,vanFrankSciRep2016,OmranScience2019} thanks to its efficiency and versatility. 

\begin{figure}[t!]
   \centering
   \includegraphics[width=145mm]{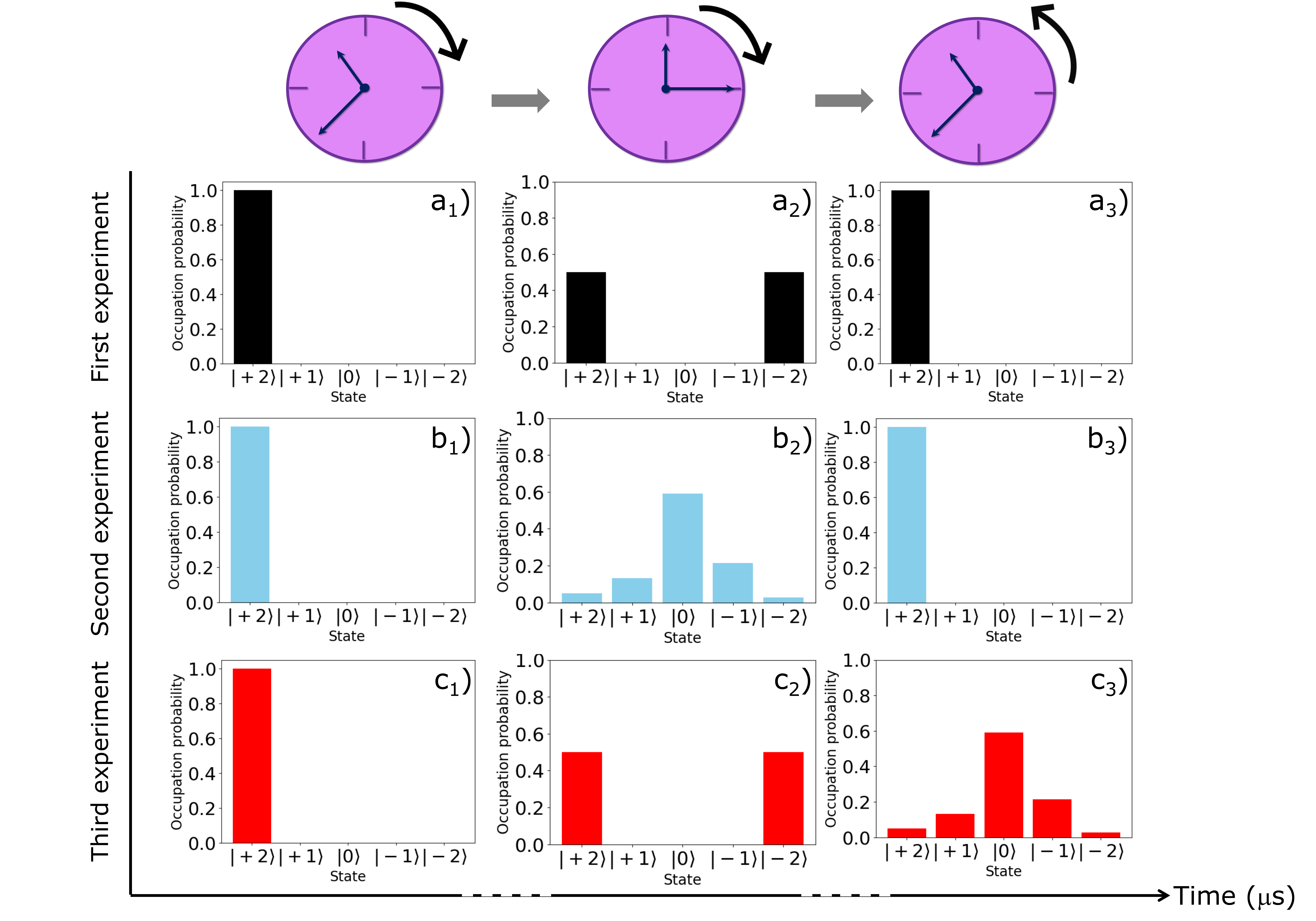}
   \caption{
   \textbf{Pictorial representation of the realized time-reversal experiments.} We prepare a $^{87}$Rb BEC in $|F=2, m_F=+2\rangle$ as the initial state for all the performed experiments denoted with a), b) and c). In the first experiment, the atoms evolve from a$_1)$ towards the state a$_2)$ where their population is equally distributed among the states $|F=2, m_F=+2\rangle$ and $|F=2, m_F=-2\rangle$, and then brought back to the initial state $|F=2, m_F=+2\rangle$ corresponding to the configuration a$_3)$. In the second experiment, the atoms evolve from b$_1)$ towards a state b$_2)$ by using the optimal pulse employed in the first experiment but with a shorter length that belongs to $[10,100)~\mus$. Then, the atoms evolve back to the initial state -- configuration b$_3)$ -- as in the first experiment. Finally, in the third experiment, the forward evolution of the atoms from c$_1)$ to c$_2)$ is the same of that in the first experiment, while in the backward process the system reaches the quantum state c$_3)$ that has been already explored in the second experiment (thus, in its past). It is worth noting that also the latter transformation realizes a quantum \emph{undo} operation, but on a shorter time-scale with respect to the other cases illustrated in the figure. Although for illustrative purposes we have chosen to report in this plot only the populations of the quantum states, in Sec.\,\ref{subsec:Exp3} we will show that the state c$_3)$ has the same (within experimental error) quantum coherence terms of b$_2)$.
   }
   \label{Pictorial}
\end{figure}

In this paper, we theoretically and experimentally exploit the dCRAB OC techniques to successfully perform time-reversal transformations, thus inverting the dynamical evolution of a quantum system realized with ultra-cold atoms from an atom-chip device, as summarized in Fig.\,\ref{Pictorial}.
To make an illustrative comparison, we also show the large differences between the results from our experiments and the ones given by inverting, through the addition of a phase term (i.e., a pre-factor $e^{i\pi}$), the time-dependence of the external driving field $f(t)$ used to address the atoms. In fact, in the absence of decoherence as in our case (at least until around $100~\mus$ of the system dynamics), the evolution of a driven quantum system is governed by the Schr\"{o}dinger equation \cite{dirac} in which the Hamiltonian operator is generally composed by two distinct contributions. One describes the inner structure of the system (the atomic Hamiltonian $H_0$), while a control term $H_{RF}$, as detailed in Sec.\,\ref{sec:physical_system}, models the action of an external time-dependent coherent field $f(t)$ that steers its dynamics. Hence, leading the system back to its initial state is not simply yielded by the inversion of the time of the driving field, i.e., $f(-t)$, because of the unavoidable presence of the Hamiltonian $H_0$ that \emph{always} evolves forward in time. It has been already proven that, in some specific cases, it is possible to exploit the periodicity of the quantum dynamics to retrace part of the evolution or to create an echo of the initial state \cite{NMR,echo1,echo2}. Some peculiar time inversion tasks have been demonstrated, such as the reversal of atom-field interaction in a cavity quantum electrodynamics experiment \cite{haroche} or feedback-control-based deterministic reversal of projective measurements on a trapped ion experiment through a quantum error-correction protocol \cite{blatt}. However, these strategies may be viable if no constraints on the duration of the time-reversal transformations are taken into account. For example, in our case using ultra-cold atoms within an atom chip device -- but similarly even in many other atomic, molecular, condensed matter and optical systems -- any dynamical transformation is constrained by the decoherence time $T_2$ \cite{FootBook}, which defines the period after which, on average, the system looses quantum coherence due to the presence of an external field and/or the coupling to the environment. In our experiments, for instance, the quantum system dynamics cannot be longer than around $100~\mus$. Up to this duration, indeed, the effects of decoherence can be almost neglected.

The paper is organized as follows. In Sec.\,II we introduce both the experimental setup and the quantum system Hamiltonian, and then explain how the driving field is optimally designed by means of the dCRAB algorithm. In Sec.\,III, instead, we present all the experiments we realized to test time-reversal transformations with ultra-cold atoms, as a proof-of-principle of \emph{undo} operations in quantum regimes. Finally, Secs.\,IV and V conclude the paper, by discussing the relevance of our experimental results and also providing a thermodynamic interpretation whereby the employed optimal control strategy corresponds to an entropy rectification procedure.

\section{Physical system and optimization protocol}\label{sec:physical_system}

\begin{figure}[h!]
   \centering
   \includegraphics[width=140mm]{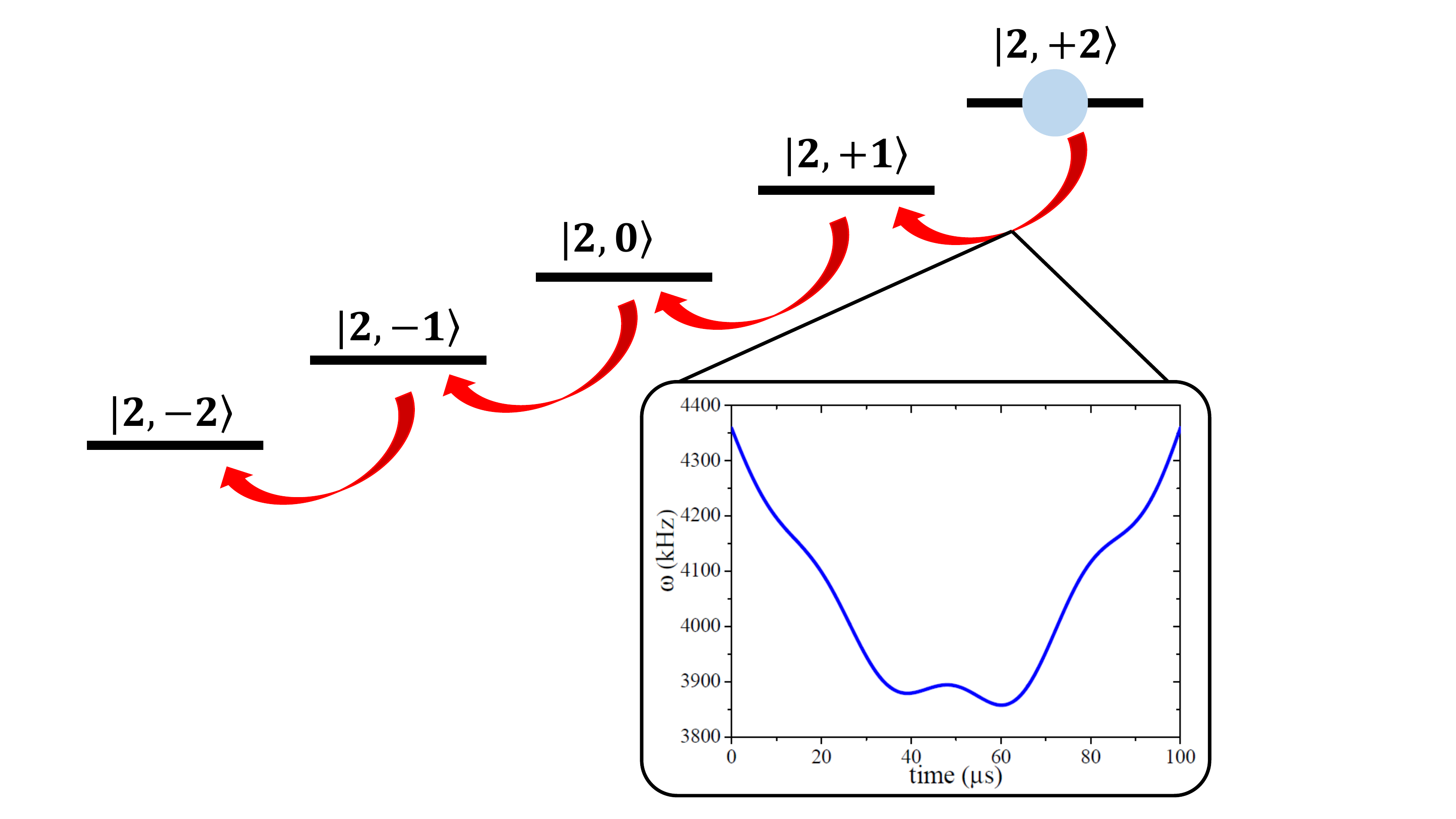}
   \caption{\textbf{$^{87}$Rb atomic energy levels.} The quantum dynamics in our time-reversal experiments take place in the $F=2$ hyperfine ground state of a $^{87}$Rb BEC. The manifold is given by the five possible orientations of a spin-$2$, energetically separated by means of a homogeneous magnetic field. The atomic cloud is initially prepared in the $|F=2,m_F=+2\rangle$ quantum state, and subsequently the five neighboring $|F,m_F\rangle$ states are coupled by a quasi-resonant radio frequency radiation (depicted by the red arrows). By modulating the latter in time through an optimally designed strategy, the energies of the five sub-levels are effectively ``shaken'' in order to drive the system back and forth in time. The inset shows an example of optimally prepared pulse, whose frequency $\omega(t)$ typically belongs to the range $1\div10 \rm\,MHz$.}
   \label{Rb_levels}
\end{figure}

The experiment is performed on a BEC of $^{87}$Rb realized with an atom chip evolving on the five-fold Hilbert space given by the $F=2$ rubidium hyperfine ground state (see Appendix for technical details). Hence, we assume that the internal state of the atomic system is described at each time $t$ by the $5 \times 5$ density matrix $\rho(t)$ in the $|F, m_F\rangle$ basis.  After the creation of the BEC, at the beginning of the system evolution, the atoms are optically pumped in the $|F=2, m_F=2\rangle$ sub-level, as shown in Fig.\,\ref{Rb_levels}.  
The free evolution of the BEC atoms is governed by the time-independent atomic Hamiltonian $H_0$ that is evaluated via the Breit-Rabi formula, which quantitatively determines the energies of all different sub-levels for a known magnetic field intensity \cite{Bransden}. In particular, for our system the atoms are subject to a constant bias magnetic field that we set to $6.179 \rm\,G$ (again, see Appendix for details). As a result, we obtain the atomic Hamiltonian $H_0 = 2\pi\hbar \ \text{diag}(8635,4320,0,-4326,-8657)\rm\,kHz$, where $\hbar$ is the reduced Planck constant and the elements of the state basis are chosen to correspond to the hyperfyne levels from $m_F = +2$ to $m_F = -2$ by ensuring that the reference zero-energy state is $|F=2,m_F=0\rangle$ (see Fig.\,\ref{Rb_levels}). 

\begin{figure}[t!]
   \centering
   \includegraphics[width=95mm]{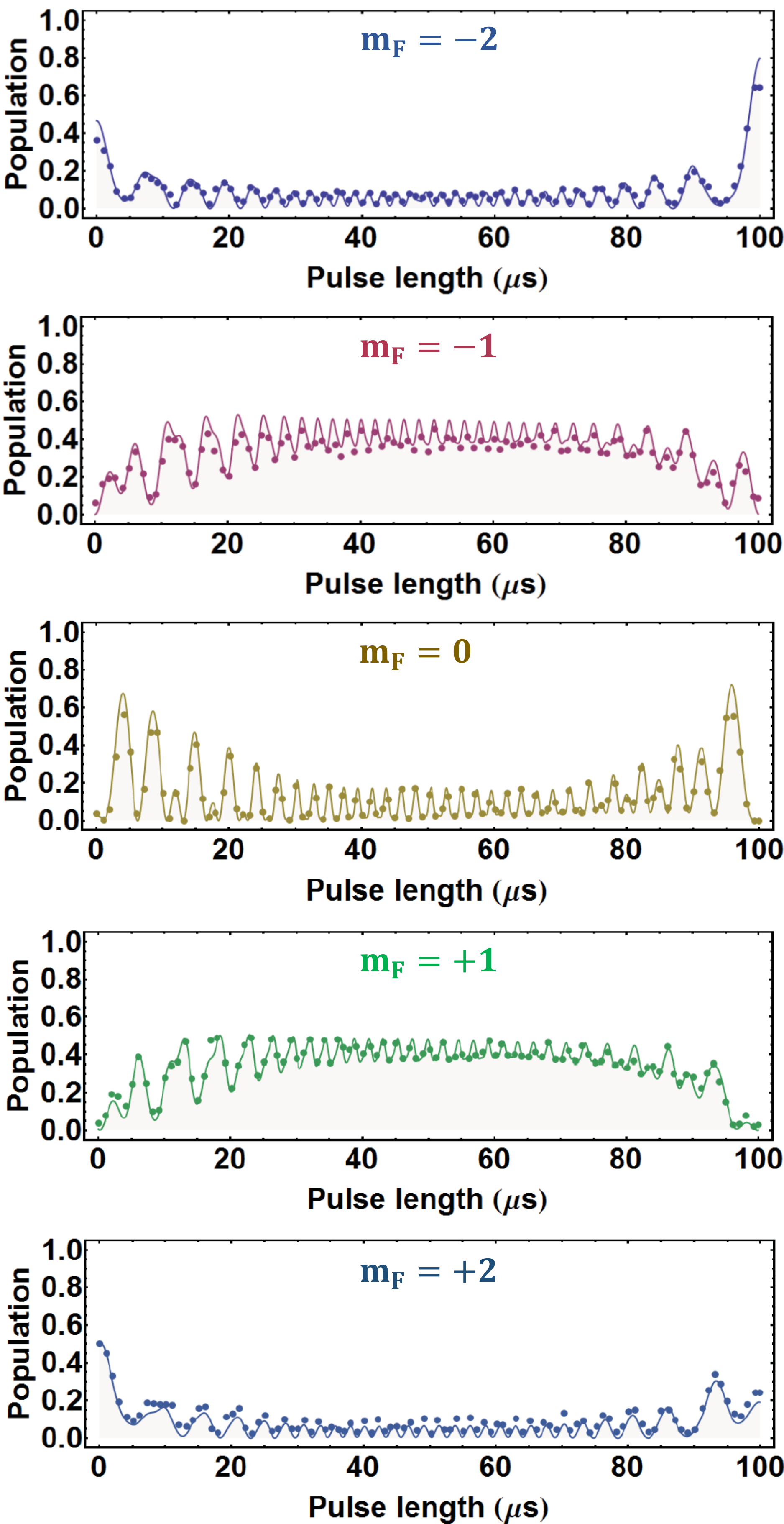}
   \caption{\textbf{Illustration of atomic population dynamics.} Time evolution of the five $m_F$ sub-levels, as an effect of the application of the optimal control pulse $f(t)$. The latter drives the atoms within the BEC from an initial state, where the population is equally distributed among the states $\vert F=2; m_F=+2 \rangle$ and $\vert F=2; m_F=-2 \rangle$, to a final state where all the population occupies the $\vert F=2; m_F=-2 \rangle$ sub-level. Continuous lines represent the results from the theoretical simulations obtained by determining the optimal quantum evolution of the system. Dots are the experimental values measured by averaging over $10$ experimental repetitions. Error bars, computed via standard deviation, are smaller then the diameter of the points and thus not shown.}
    \label{Populations}
\end{figure}
The atomic evolution is driven through a quasi resonant radio frequency (RF) field, which is produced by micro structured conductors integrated on the atom chip. Specifically, the driving is implemented through a frequency-modulated RF pulse $f(t)$ that couples the five neighboring $m_F$ states described by the following Hamiltonian, expressed in the $|F, m_F\rangle$ basis and valid in the rotating wave approximation (RWA) regime:
\begin{equation}
H_{RF}(t)=\hbar
    \left(\begin{array}{ccccc}
        -2 f(t) & \Omega & 0 & 0 & 0 \\
        \Omega & -f(t) & \sqrt{3/2} \ \Omega & 0 & 0 \\
        0 & \sqrt{3/2} \ \Omega & 0 & \sqrt{3/2} \ \Omega & 0 \\
        0 & 0 & \sqrt{3/2} \ \Omega & f(t) & \Omega \\
        0 & 0 & 0 & \Omega & 2f(t)
    \end{array}
    \right),
    \label{driving}
\end{equation}
where $f(t) = \partial_t[t\omega(t)]$ and $\omega(t)$ denotes the time-dependent frequency of the driving field. The coupling Rabi frequency $\Omega$ is proportional to the RF field amplitude and in the following will be set to $\Omega = 2\pi\times60.0~\kHz$. 

Overall, the total Hamiltonian describing the system is $H(t) = H_0+H_{RF}(t)$. Moreover, we may also include in the model a dephasing term by means of super-operators expressed in the Gorini-Kossakowski-Sudarshan-Lindblad (GKSL) form to describe the presence of experimental low-frequency noise on the magnetic bias field and on the RF signal \cite{PetruccioneBook}. In order to drive the system evolution back and forth in time, the optimal time dependence of $f(t)$ needs to be determined. This goal is achieved by minimizing the difference between the final and target quantum states of the atomic evolution, both expressed in terms of the density matrix $\rho$. This difference is provided by the error function $\epsilon \equiv \frac{1}{2} \sum_{n=1}^{5} \left|\rho_{n,n}(T) - \widehat{\rho}_{n,n}\right|$, where $T$ is the duration of the control pulse, $\rho_{n,n}(T)$ denotes the final atomic population of the $n$-th sub-level at $t=T$, while $\widehat{\rho}_{n,n}$ is the corresponding target population.

To minimize the error function $\epsilon$, the time dependence of the (slowly oscillating) frequency RF control pulse $f(t)$ is optimally modulated by following the prescriptions of the dCRAB method \cite{CRAB,DCM,CanevaPRA2014,RachPRA2015,MuellerReview2021}. For this purpose, the time-dependent frequency $\omega(t)$ of the driving field is expanded in the standard Fourier basis such that:
\begin{equation}\label{f_t}
	f(t) = 1 + \sum_{k=-7}^{7} A_k (1 + i \nu_{k} t) e^{i\nu_{k}t}\,,
\end{equation}
and the optimal values of the expansion coefficients $A_k$ (amplitude of the control function modulation) are determined by ensuring that the error function $\epsilon$ is minimized. In Eq.\,(\ref{f_t}) $\nu_k=2\pi k/T$, where $k$ is the index that spans the set of harmonics pertaining to the driving field, with $k=1,\ldots,7$ \cite{LloydPRL2014}, and $T$ denotes the length of the control pulse as above. Moreover, the optimization procedure in determining the optimal values of $A_k$ is performed via the subplex variant of the Nelder-Mead algorithm \cite{rowan}. 

The time-reversal protocol introduced in this work operates only on the diagonal elements of the final density matrix $\widehat{\rho}(T)$ (reached at the end of the evolution), corresponding to the quantum system populations that we directly measure. In fact, the optimization procedures, which we perform to design the optimal pulses that drive the quantum system dynamics, are set not to employ non-diagonal elements of $\widehat{\rho}(T)$ that should be necessarily measured by means of a tomography process. In this regard, by making use of the results in Refs.\,\cite{nostroopt,nostrotom} concerning the optimal preparation of quantum states on $^{87}$Rb BEC atom-chip-based micro-traps, we implement a preliminary test experiment to tune the values of the setup parameters (i.e., the constant magnetic field and Rabi frequency) for accurate state preparation and transfer. In this test experiment, we compare the theoretical and experimentally measured time evolutions of the atomic population in each of the five $m_F$ sub-levels during the application of an optimal pulse. The latter brings the quantum system from an initial state, in which the population is equally distributed among the states $|F=2,m_F=+2\rangle$ and $|F=2,m_F=-2\rangle$ in a coherent superposition, to a final state where all the population occupies the $|F=2,m_F=-2\rangle$ sub-level. The experimental results, reported in Fig.\,\ref{Populations} for illustrative purposes, are in satisfactory agreement with the theoretical predictions obtained by solving the Liouville-von Neumann differential equation $\dot{\rho}(t) = -(i/\hbar)[H(t),\rho(t)]$.

\section{Experiments}\label{sec:Experiments}

\begin{figure}[htpb]
\centering
\begin{subfigure}[b]{0.3\textwidth}
\includegraphics[width=\textwidth]{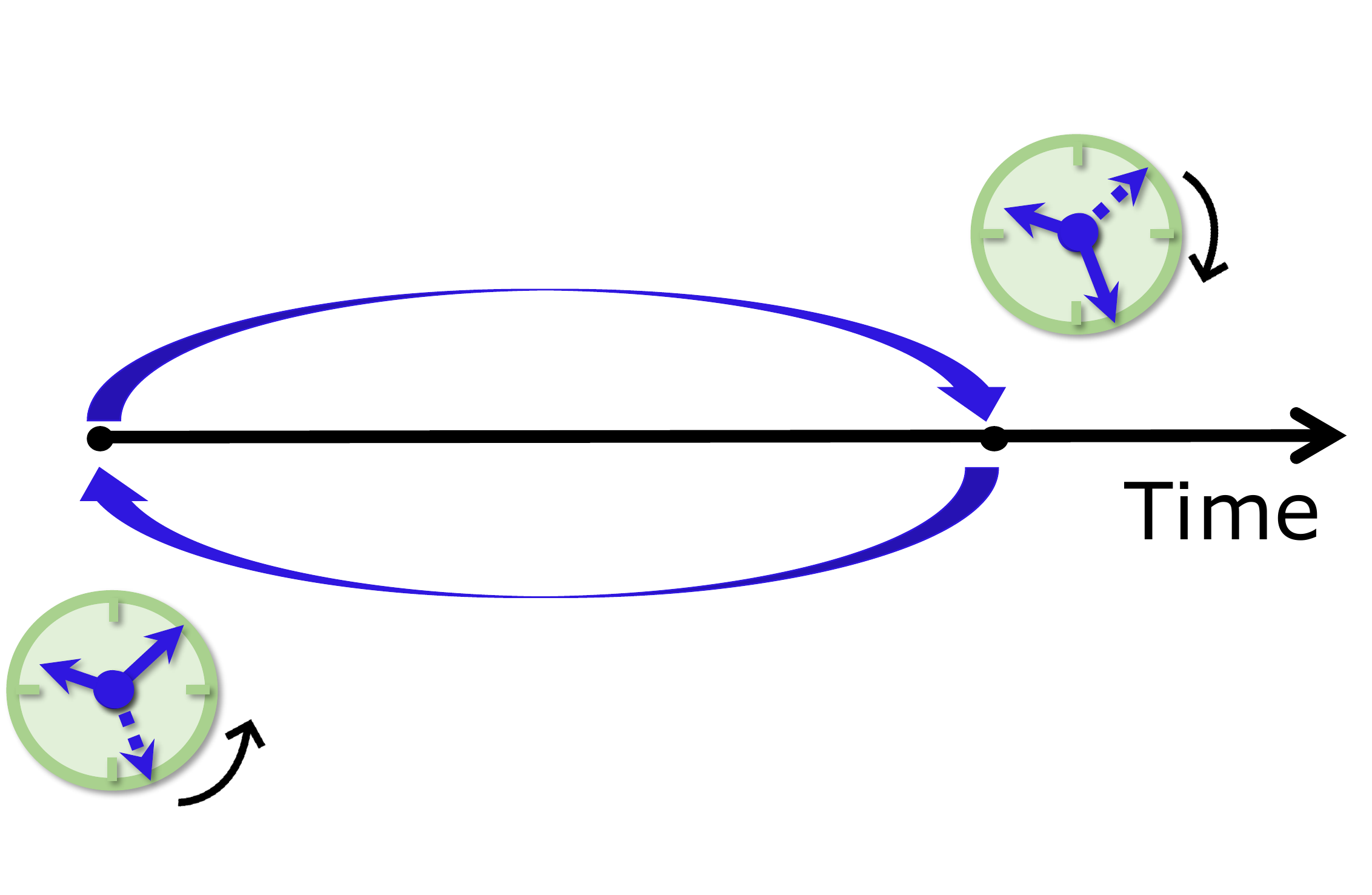}
\caption{\footnotesize{First set of experiments}}
\label{subfig:Exp-I}
\end{subfigure}
~
\begin{subfigure}[b]{0.3\textwidth}
\includegraphics[width=\textwidth]{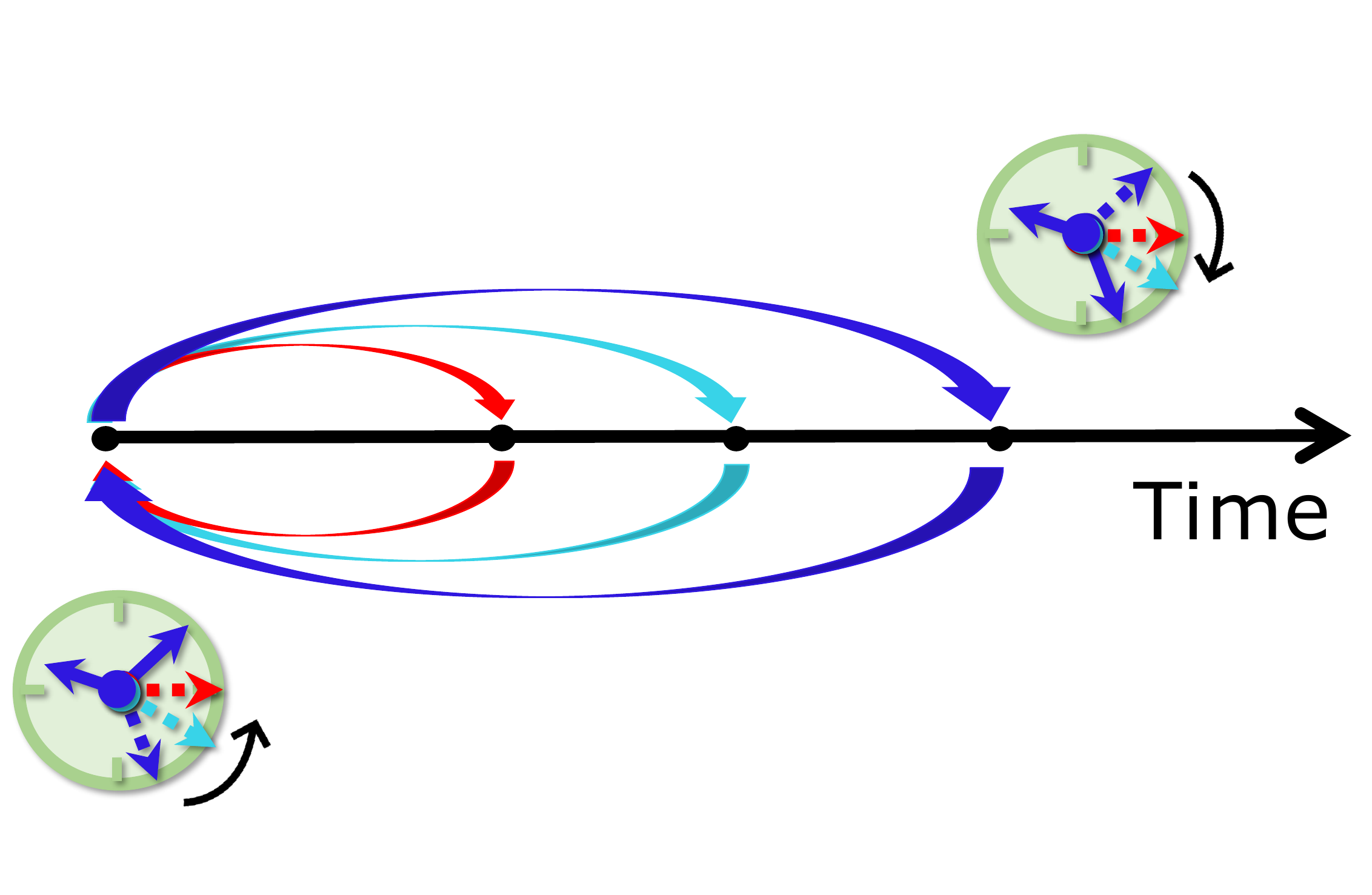}
\caption{\footnotesize{Second set of experiments}}
\label{subfig:Exp-II}
\end{subfigure}
~
\begin{subfigure}[b]{0.3\textwidth}
\includegraphics[width=\textwidth]{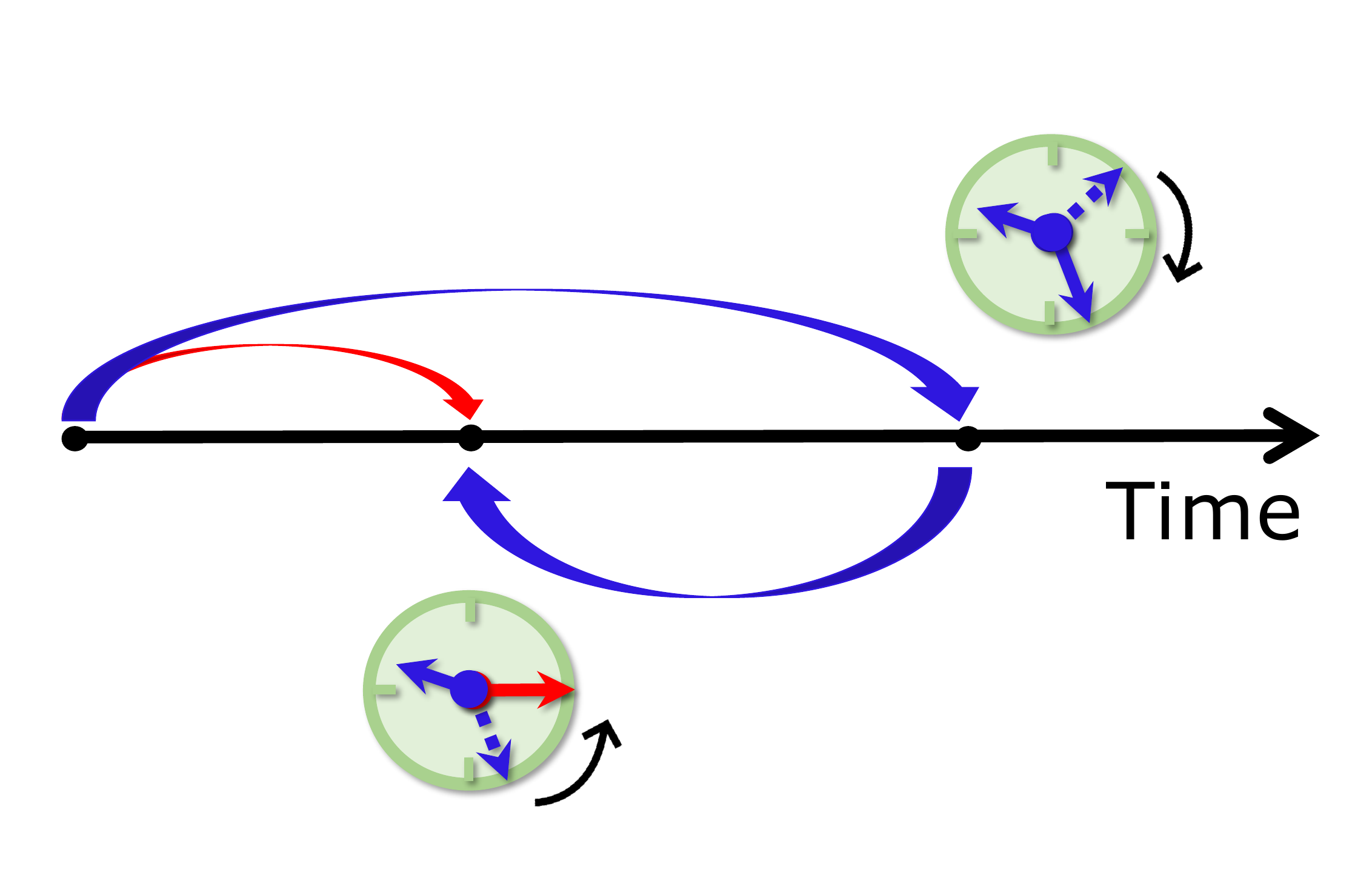}
\caption{\footnotesize{Third set of experiments}}
\label{subfig:Exp-III}
\end{subfigure}
~
\caption{\textbf{Experiments.} (a) The atomic evolution is driven forward and backward between an initial state and a given target one. (b) The time-reversal of the quantum system evolution is performed along trajectories with gradually shorter time duration. (c) The quantum state of our system is driven back in time to a quantum state that has been already explored in the past.}
\label{fig:PicExperiments}
\end{figure}

In this work we exploit the dCRAB control techniques to realize three different set of experiments (see Fig.\,\ref{fig:PicExperiments}) to faithfully time-invert the evolution of a quantum system realized with ultra-cold atoms.
Starting from the initial state where the atomic population occupies the $|F=2, m_F=2\rangle \equiv |+2\rangle$ level (as depicted in Fig.\,\ref{Rb_levels}), the proposed strategies are successfully applied along several paths in the Hilbert space of the system. Here, we perform the time-inversion of quantum operations by using gradually higher levels of control in terms of the complexity of the addressed control problems. Moreover, we are also going to illustrate how such techniques allow for the extension of the implemented time-reversal transformations to much more complicated situations, in which performing the backward evolution in the shortest time-scale and/or with accuracy values as high as possible may be crucial. Finally, in order to make a comparison and demonstrate the need to employ quantum OC methods, we show how different the outcomes of such experiments are if one inverts the time-dependence of the external driving field instead of using an optimal driving pulse.
In doing this, whenever a full density matrix reconstruction is performed for additional test experiments (see the next sections), the distance between the target and the experimentally measured quantum states, $\widehat{\rho}$ and $\rho(t)$ respectively, is evaluated through the Uhlmann fidelity \cite{Uhlmann1976}
\begin{equation}
    \mathfrak{F}(\widehat{\rho},\rho(t)) \equiv \left( {\rm Tr}\sqrt{\sqrt{\widehat{\rho}}\rho(t)\sqrt{\widehat{\rho}}} \right)^{2}.
\end{equation}
In all the other cases, the accuracy in performing a given operation is assessed by means of the error function, according to the formula $1 - \epsilon$.

\subsection{First set of experiments}\label{subsec:Exp1}

In the first set of experiments (Fig.\,\ref{subfig:Exp-I}), our aim is to drive the evolution of the quantum system forward and backward: from an initial state $\rho(0)$ to a given target one $\widehat{\rho}$ and then back again to $\rho(0)$. This experiment is performed twice: firstly by time-inverting the driving field $f(-t)$, and secondly by controlling the time-reversed evolution via the optimally-designed driving pulse $f_{OC}(t)$, where again the subscript OC stands for `optimal control'. The results obtained in both cases are then compared.

\begin{figure}[t!]
   \centering
   \includegraphics[width=185mm]{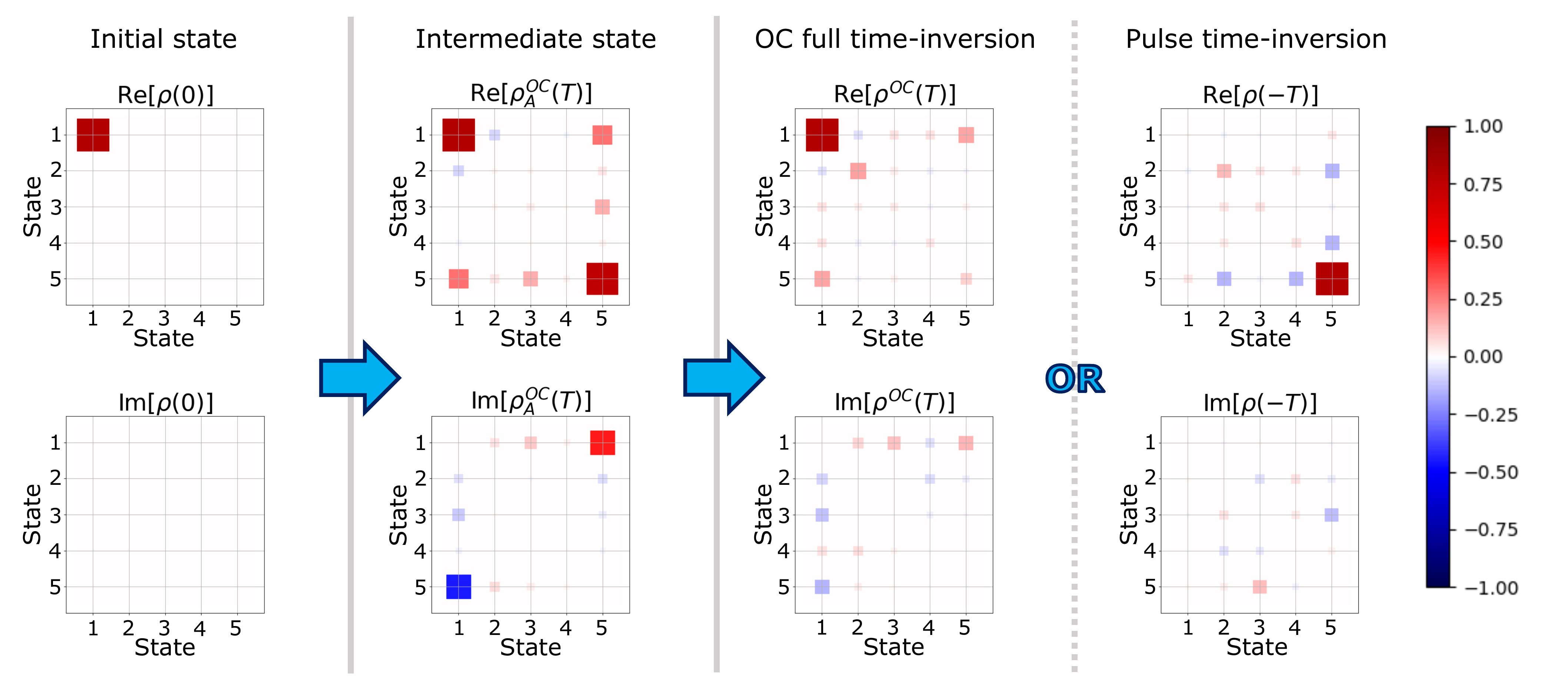}
   \caption{\textbf{Tomographic reconstruction.} Density matrix representation in Hinton plots (i.e., diagrams for visualizing the numerical values of the elements composing a matrix) of the initial state $\rho(0)$, the intermediate state $\rho^{OC}_A(T)$ (reached via OC and as much close as possible to the target state $\widehat{\rho}_A$), and the final states $\rho^{OC}(T)$ and $\rho(-T)$ obtained by inverting, respectively, the whole system quantum dynamics via OC techniques or the pulse time-dependence of the driving field. The positive and negative numerical values of the matrices elements are here represented by red and blue squares respectively, while their magnitude is directly proportional to the size of the depicted squares.}
   \label{Tomo_reconstruction}
\end{figure}
The experiment is repeated four times to test the realization of an accurate time-reversal transformation over four different paths in the Hilbert space of the BEC. Specifically, our quantum system is driven -- according to the optimal strategy of Ref.\,\cite{nostroopt} -- from the initial state $\rho(0)$, such that $\rho_{1,1}(0) = 1$ and $\rho_{k,j}(0) = 0$ for $k,j=1,\ldots,5$ apart $k=j=1$, to the following four different target states: i) $\widehat{\rho}_A$: $\widehat{\rho}_{1,1} = \widehat{\rho}_{5,5} = 0.5$; ii) $\widehat{\rho}_B$: $\widehat{\rho}_{2,2}=\widehat{\rho}_{4,4}=0.5$; iii) $\widehat{\rho}_C$: $\widehat{\rho}_{1,1}=\widehat{\rho}_{2,2}=0.5$; iv) $\widehat{\rho}_D$: $\widehat{\rho}_{n,n}=1/5$ for $n=1,\ldots,5$, where for each target all the other elements of $\rho$ are equal to zero. Subsequently, the optimal control pulse $f_{OC}(T)$ (designed as in Sec.\,\ref{sec:physical_system}) or $f(-T)$ are applied to the BEC to bring the system back to the initial state $\rho(0)$. In all the analyzed cases, the time duration $T$ of the forward and backward processes is set to $100~\mus$, thus entailing a total system evolution of $200~\mus$. 

It is worth noting that our results are validated by using just the atomic populations of the system, under the assumption of unitary dynamics. However, it is well known that measuring the population elements of a quantum system represents only a partial knowledge of its full density matrix. For this reason, despite the good agreement between theoretical continuous lines and experimental dots in Fig.\,\ref{Populations} that seems to confirm our assumption of unitary dynamics, we perform a full density matrix reconstruction for the case i). As reported in Fig.\,\ref{Tomo_reconstruction}, we have measured the density matrix of the experimental state $\rho^{OC}_A(T)$ that is reached in the forward evolution by following the optimized OC path from the initial state $\rho(0) = |+2\rangle\!\langle +2|$ to the intermediate target state $\widehat{\rho}_A$. Then, the two possible final density matrices $\rho^{OC}(T)$ and $\rho(-T)$, corresponding respectively to the ending stage of the optimally controlled and time-inverted backward trajectories, are reconstructed. The results illustrated in Fig.\,\ref{Tomo_reconstruction} show that inverting only the time-dependence of the driving field brings the BEC atomic population closer (in the sense given by the Uhlmann fidelity) to the orthogonal state $|-2\rangle$ instead of $|+2\rangle$, while the optimally reversed evolution successfully reaches the initial state.
\begin{figure}[t!]
   \centering
   \includegraphics[width=125mm]{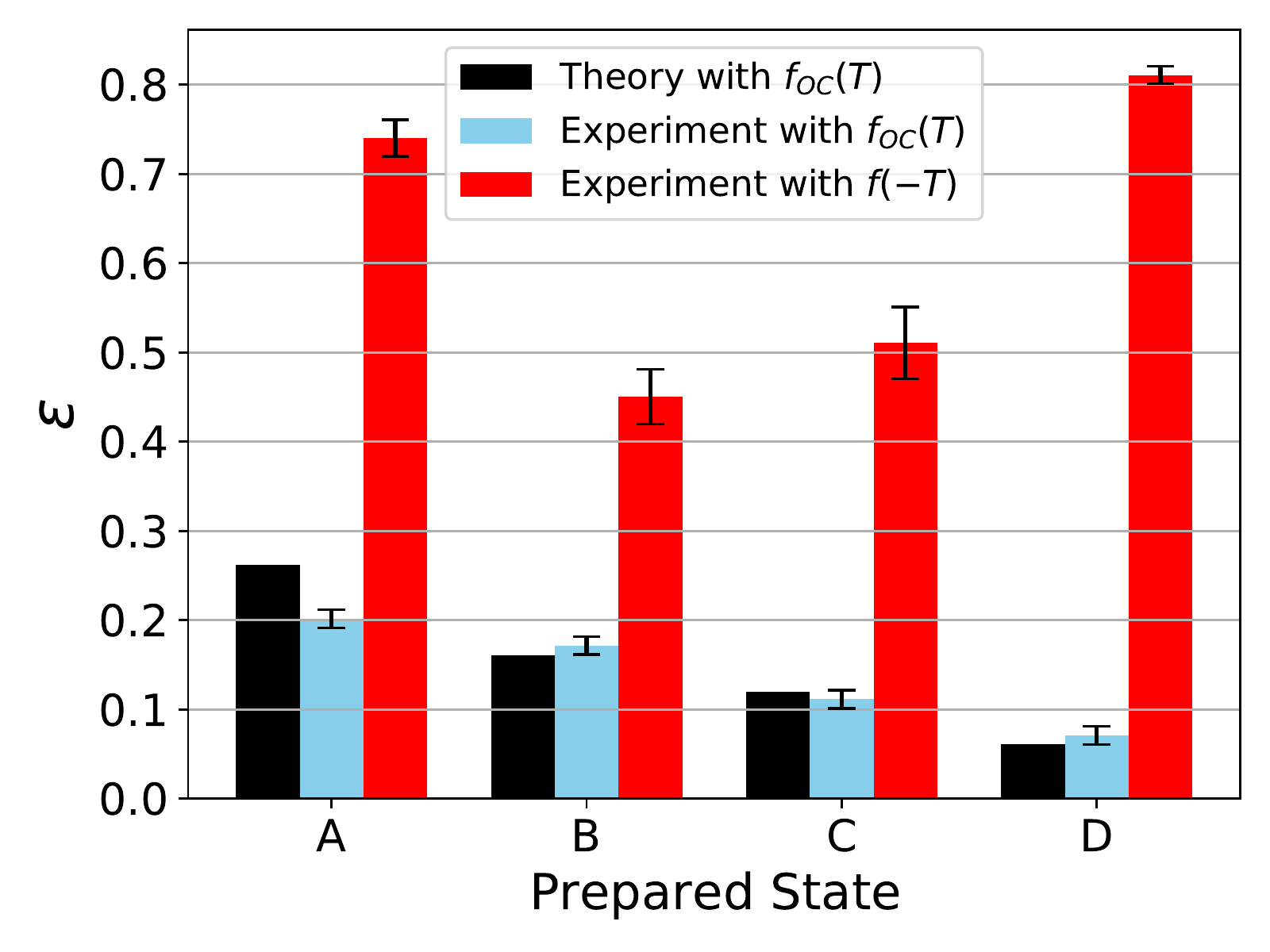}
   \caption{\textbf{Control error functions.} Theoretical and experimental error function computed for the implemented time-reversed quantum dynamics in reaching the initial state $\rho(0)$. Here, we start from the target states $\widehat{\rho}_A$, $\widehat{\rho}_B$, $\widehat{\rho}_C$, $\widehat{\rho}_D$ within the achievable experimental accuracy. The black bars represent the numerically simulated error function $\epsilon$ (defined in Sec.\,\ref{sec:physical_system}) obtained through the optimized inversion of the driving field $f_{OC}(T)$. The light blue bars are the corresponding experimental error function (with its standard deviation), while the red bars denote the experimental error (with its standard deviation) obtained by changing the time-dependence of the driving field and thus applying the pulse $f(-T)$. The error bars are computed by repeating $10$ times each set of experiments. It is worth noting that the error function is never equal to zero even in our numerical simulations. This is due firstly to the presence of decoherence, as we will show in Sec.\,\ref{subsec:Exp2}, and secondly to the limited number of resources at disposal (in terms of operations) to carry out the OC protocol.}
   \label{Error_function}
\end{figure}
Finally, we have implemented the forward and backward evolution for all the remaining paths ii), iii) and iv). The resulting values of the error functions $\epsilon$ are illustrated in Fig.\,\ref{Error_function} where similar behaviors can be observed for all the tested target states i), ii), iii) and iv).

\subsection{Second set of experiments}\label{subsec:Exp2}

To evaluate the time limits/constraints of the optimally-controlled time-reversal transformations implemented in the first set of experiments, we perform a second set of experiments (Fig.\,\ref{subfig:Exp-II}) to drive the quantum system evolution back and forth from the state $\rho(0)$ to quantum target states $\widehat{\rho}_{Q_j}$, along the same trajectory, by using pulses of gradually shorter lengths $T_j$, which belong to the set $\{10, 20, 40, 60, 70, 80, 100\}~\mus$. In more detail, first we design an optimal forward pulse that brings the quantum system from the initial state $\rho(0)$ to the state $\widehat{\rho}_A$ reached at $T=100~\mus$. Then, we interrupt this pulse at each $T_j$ in order to obtain the other sub-pulses that realize gradually shorter evolutions. In all these experiments, the different pulses that realize the forward and backward processes have the same duration. The backward process from each $\widehat{\rho}_{Q_j}$ to the quantum state $\rho(0)$ is realized, once by inverting the time-dependence of the forward pulse and another time using OC to design the backward pulse, similarly to what done in the first set of experiments. 
\begin{figure}[t!]
   \centering
   \includegraphics[width=125mm]{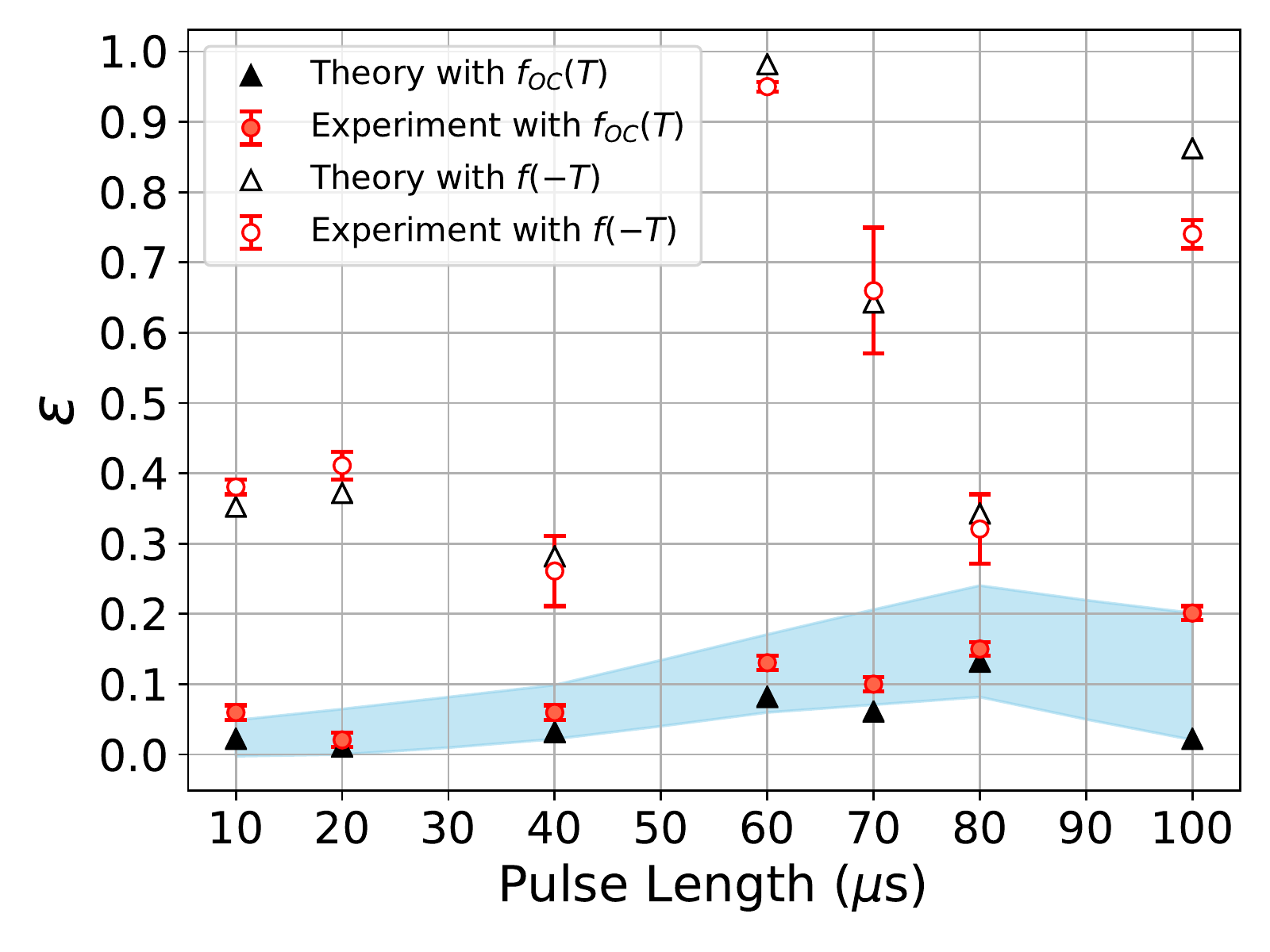}
   \caption{{\bf Testing optimal time-reversal over time.} Error function vs pulse length $T$, with $T\in\{10, 20, 40, 60, 70, 80, 100\}~\mus$, from both the theoretical and experimental side. The values of the error function are evaluated, once by using the proposed optimal control strategy and another by inverting the time-dependence of the driving field. In the figure, we also take into account the presence of experimental dephasing noise on the quantum system evolution, by including in the numerical simulations a correction term that adjusts the theoretical prediction. The correction, pictorially represented in the figure by the light-blue shaded area, is numerically simulated for each value of $T$ by solving the Liouville-von Neumann differential equation in the dephasing range $\gamma_n \equiv \gamma \in 2\pi[20, 200]\rm\,Hz$ (estimated at $T=100~\mus$) and with an additional magnetic field fluctuation with standard deviation of $1~mG$. In this way, the correction is finally obtained by taking the corresponding minimum and maximum values of such a computation as explained in the main text.}
   \label{Time_dependence}
\end{figure}
The experimental results reported in Fig.\,\ref{Time_dependence} show a smaller error $\epsilon$ in realizing reversed quantum dynamics via OC techniques compared to the ones obtained by changing the time-dependence of the driving field, thus confirming the results found in Subsec.\,\ref{subsec:Exp1} but over quantum dynamics with a shorter time-scale. Furthermore, by accounting in the theoretical model for experimental dephasing noise that entails quantum coherence degradation, the experimental results are in good agreement with the corresponding theoretical predictions for driving pulses with not so long duration, while for experiments longer than $80~\mus$ the mismatch slightly increases. These effects on the error function are depicted by the light-blue shaded area in Fig.\,\ref{Time_dependence}. In more detail, dephasing noise is included in the model by means of the following Lindbladian super-operator term ${\cal L}$ acting on the density matrix $\rho(t)$: ${\cal L}(\rho(t)) = \sum_{n=1}^{5} \gamma_n \left[- \left\{|n\rangle\!\langle n|,\rho(t)\right\} + 2|n\rangle\!\langle n|\rho(t)|n\rangle\!\langle n|\right]$, where $\{\cdot,\cdot\}$ denotes the anti-commutator and $\gamma_n$ are the dephasing rates. The action of ${\cal L}$ is to randomize the phase of each sub-level $n$ of the BEC with rate $\gamma_n$. Hence, the light-blue shaded area is obtained as follows. The difference between the experimental and theoretical points at $100~\mus$ is attributed exclusively to the dephasing noise that determines the range of $\gamma$. Then, starting from such dephasing range, the lower and upper bound of the shaded area at each pulse length $T$ are obtained by numerically solving the GKSL equation: $\dot{\rho}(t) = - \frac{i}{\hbar} [H(t),\rho(t)] + {\cal L}(\rho(t))$, by choosing the dephasing rate $\gamma_n \equiv \gamma$ constant for all the sub-levels in the interval $2\pi[20, 200]\rm\,Hz$ and considering magnetic field fluctuations within the range $\Delta B = 1~mG$. In this regard, we recall that in the GKSL equation, which models the time evolution of the system's density matrix affected by dephasing noise, $H(t) \equiv H_0 + H_{RF}(t)$ with $H_{RF}(t)$ defined as in Sec.\,II and $f(t)$ is constrained in the range $f(t) \in 2\pi [4150, 4600]~\kHz$, so as to maintain always the same coupling of the RF antenna to the driving circuit.

\subsection{Third set of experiments}\label{subsec:Exp3}

To better illustrate the wide applicability of our implemented time-reversal procedures, in a third set of experiments (see Fig.\,\ref{subfig:Exp-III}) we aim to show that our OC strategy is able to invert the evolution of the quantum system by driving it back from $\widehat{\rho}_A$ (target state at $T=100~\mus$) to a quantum state $\rho_{P}(\tau)$ that is reached in $\tau \leq 100~\mus$ along the same trajectory linking $\rho(0)$ with $\widehat{\rho}_A$. In this way, we are going to show that it is also possible to drive the system back to a generic quantum state that has been already explored in the past. This effectively qualifies our experiments as a proof-of-principle of quantum \emph{undo} operations, whereby the external user has to be able to reverse at will the last operation they performed.

\begin{figure}[t!]
   \centering
   \includegraphics[width=125mm]{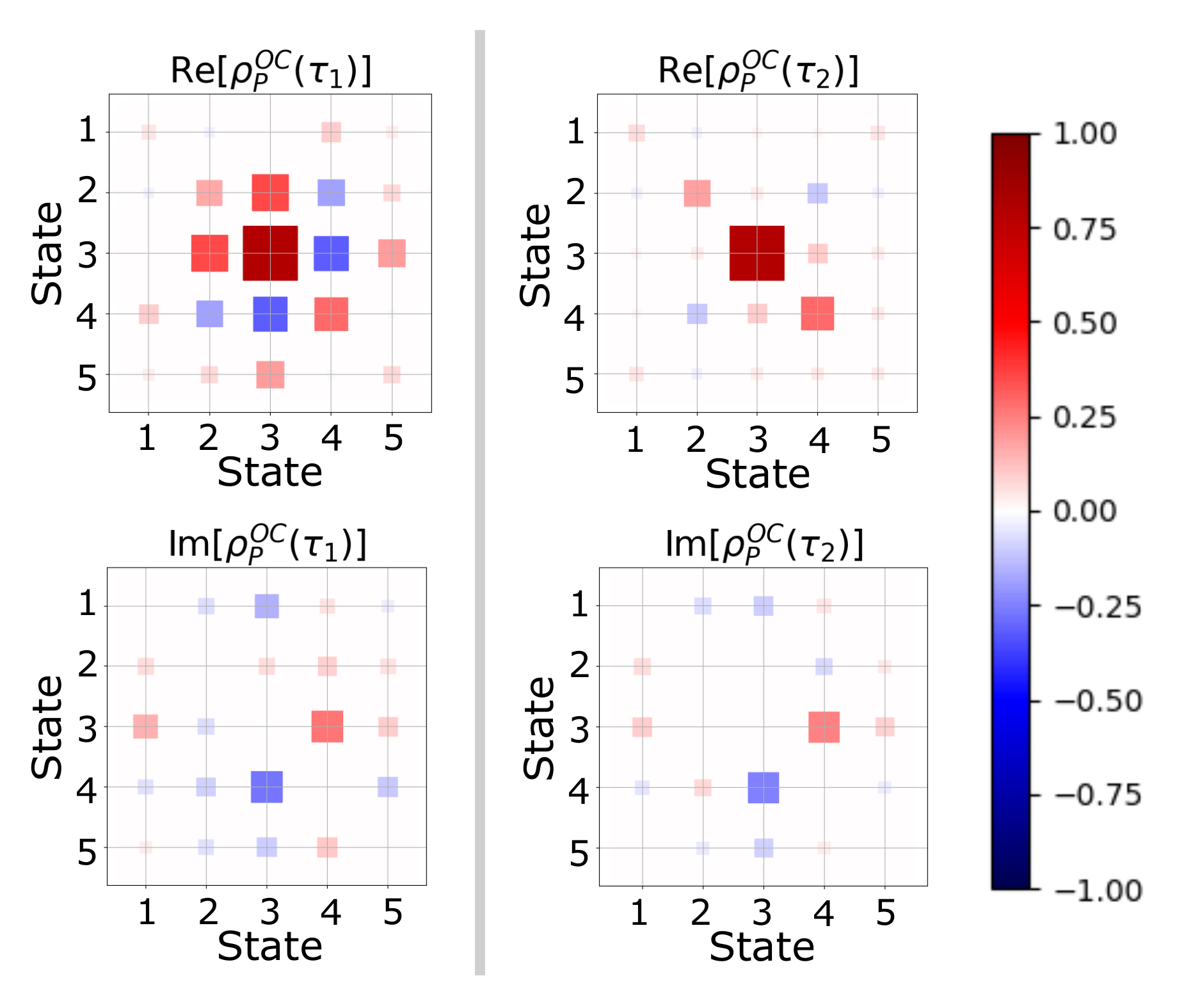}
   \caption{{\bf Tomography reconstruction of quantum states $\rho^{OC}_P(\tau_1)$ and $\rho^{OC}_P(\tau_2)$.} The state $\rho^{OC}_P(\tau_1)$, reported on the left-hand side of the figure, is reached in $\tau_1=33~\mus$ in the first stage of the experiment and represents a state in the past of the quantum system evolution. Instead, the state $\rho^{OC}_P(\tau_2)$ on the right-hand side of the figure is reached in $\tau_2=67~\mus$ via the optimal time-reversal procedure we have realized. The accuracy 1-$\epsilon$ between these two measured density matrices is about $97.3\%$.}
   \label{fig:Hinton_exp3}
\end{figure}

In particular, also in this case, the set of experiments is performed twice. First, starting from the initial state $\rho(0)$, the system evolution, enabled by the driving pulse that drives the quantum system from $\rho(0)$ to $\widehat{\rho}_A$ in $100~\mus$ (it is the same optimal pulse used in the first set of experiments), is interrupted after $\tau_{1}=33~\mus$. In that instant $\tau_1$, the system has reached the intermediate state $\rho^{OC}_P(\tau_1)$, which is then reconstructed via tomography. Therefore, this first stage of the experiment allows us to identify the state $\rho_P(\tau_1)$. Secondly, the system is made to evolve from the initial state $\rho(0)$ to the target state $\widehat{\rho}_A$ in $100~\mus$ without interrupting the pulse. Exploiting the dCRAB optimization procedure, then, a path from $\widehat{\rho}_A$ to the state $\rho_P(\tau_1)$ is traced back by using an optimal pulse lasting $\tau_{2}=67~\mus$ (note that, by construction, $\tau_{1}+\tau_{2}=100~\mus$), and the resulting state $\rho_P^{OC}(\tau_{2})$ is measured again via a tomographic process. The experimental results are reported in Fig.\,\ref{fig:Hinton_exp3}.

The accuracy $1-\epsilon$ between the measured density matrix $\rho_P^{OC}(\tau_1=33~\mus)$, representing a state in the past of the system evolution, and the measured state $\rho_P^{OC}(\tau_{2}=67~\mus)$ reached via the optimally time inverted dynamics, is around $97.3\%$. These results illustrate that the implemented OC strategy allows to perform a quantum \emph{undo} operation not only of the last quantum state $\rho_A$ but also of any past state $\rho_P(\tau)$ in the occurred quantum dynamics. In other terms, one is able to bring back the system from the target state $\widehat{\rho}_A$ to the initial one $\rho(0)$, but even from $\widehat{\rho}_A$ to a generic state along the pathway $\rho(0) \longleftrightarrow \widehat{\rho}_A$.

\section{Discussion}\label{sec:Discussion}

The introduction of a procedure to achieve time-reversal transformations is implicitly linked with the understanding of a clever way to nullify (or even rectify) the thermodynamic entropy originated by the system \cite{mackey,CamatiPRL2016,ManzanoPRX2018,GherardiniQST2018_Entropy,BrunelliPRL2018,BatalhaoBookChapter,KwonPRX2019,LandiRMP2021}. The principles of thermodynamics, and in particular the second law, tell us that if a dynamical process (classical or quantum) is reversible, then it operates to come back to the starting point in a recurrent way without further consumption of resources. For such dynamics, thus, it might not be required to carry out time-reversal procedures by means of an external drive. Clearly, this is not the case in our experiments. In fact, the decoherence time (intrinsic to the Bose-Einstein condensate) imposes as upper bound a time window of around $100~\mus$ to the experimental implementation of the dynamics, as discussed in Sec.\,\ref{subsec:Exp2}. This temporal constraint makes the quantum process that we can physically implement as it was irreversible, since the possibility that the quantum system autonomously (i.e., without the driving) comes back to the initial state is generally prevented. As a consequence, one needs to pump energy from the outside (in our case making use of OC strategies) to successfully carry out time-reversal transformations.

In our experiments, we have quantified this aspect by computing the Loschmidt echo \cite{VanicekPRE2006,GoussevScholarpedia,GoussevPTAMPES}:
\begin{equation}
M(\tau) \equiv \left|\langle\psi_0|e^{i\mathcal{H}_{2}(\tau)\tau}e^{-i\mathcal{H}_{1}(\tau)T}|\psi_{0}\rangle\right|^{2},
\end{equation}
where $\hbar$ is here set to $1$, $|\psi_0\rangle$ denotes the initial wave-function such that $\rho(0) = |\psi_0\rangle\!\langle\psi_0|$, and $\tau$ is the duration of both the forward and backward processes. Since in our case we can identify $\mathcal{H}_{1}(t) = H(t)$ as the BEC Hamiltonian of the forward process in the implemented dynamics and $\mathcal{H}_{2}(t)=H_{OC}(t)$, the Loschmidt echo $M(\tau)$ can be equivalently written as:
\begin{equation}
    M(\tau) = {\rm Tr}\left[e^{iH_{OC}(\tau)\tau}\widehat{\rho}(\tau)e^{-iH_{OC}(\tau)\tau}\rho(0)\right] = {\rm Tr}\left[\rho^{OC}(\tau)\rho(0)\right]
\end{equation}
where $\widehat{\rho}(\tau)$ is the target quantum state achieved by the forward process at $t=\tau$. The Loschmidt echo $M(\tau)={\rm Tr}[\rho^{OC}(\tau)\rho(0)]$, computed experimentally for each set of tomographic data, provides the same values of the corresponding Uhlmann fidelity values $\mathfrak{F}(\rho^{OC}(\tau),\rho(0))$. This evidence, beyond providing a thermodynamic interpretation of our experimental findings, also allows us to confirm, in a quantitative way, that the time-arrow inversion $t \rightarrow -t$ of the time-dependent terms in the interaction Hamiltonian is not sufficient in general to reverse a quantum evolution and thus to implement quantum \emph{undo} operations.

\section{Conclusions}

In this paper, we have experimentally tested the effectiveness of OC methods, enabled in our case by a dCRAB technique, to carry out time-reversal transformations with an accuracy on average around $92\%$ in a BEC on an atom chip. Specifically, we have realized three sets of experiments. In the first set, the laser-cooled $^{87}$Rb atoms of the condensate are driven forward and backward in time from an initial state $\rho(0)$ to a target one and then back to $\rho(0)$. In the second set of experiments, we have shown that the adopted OC technique works with almost equal accuracy in bringing back to the initial condition any quantum target state along the same trajectory, independently on the time instant in which the target state was achieved in the forward evolution of the system. In a third set of experiments, we also demonstrate the possibility to drive the quantum system back to a generic quantum state already explored in its past dynamics. 

The realization of \emph{undo} operations of the last-performed computation executed by a quantum circuit is the primary application of our experiments. In fact, while we cannot experience time-reversal phenomena occurring spontaneously due to the unidirectionality in time of physical processes, in a digital context as a (quantum) computing device, reversing a given operation may be a feasible task. The most significant example is the \emph{undo} command that allows to reverse a calculation that has been performed in a past step of a computational routine. By resorting to OC theory, here we have experimentally proved that the time-reversal of operations is possible also in the quantum realm. We thus expect that in the next future one can realize in commercial quantum computers quantum \emph{undo} commands that represents the main technological application of our work.

\subsection*{Acknowledgements}

The authors thank Francesco Scazza for useful and insightful comments. F.C.\,was financially supported by the European Union's Horizon 2020 research and innovation programme under FET-OPEN Grant Agreement No.\,828946 (PATHOS). S.G.\,also acknowledges The Blanceflor Foundation for financial support through the project ``The theRmodynamics behInd thE meaSuremenT postulate of quantum mEchanics (TRIESTE)''. S.M. acknowledges support from the Horizon2020 program QuantERA ERA-NET Cofund in Quantum Technologies project T-NISQ, the Italian PRIN2017 and Fondazione CARIPARO, the BMBF project QRydDemo, and the INFN project QUANTUM.

\section*{Methods: Details on the experimental procedure}\label{sec:Appendix}

All the experiments reported in the main text are realized on a collection of laser-cooled rubidium ($^{87}$Rb) atoms prepared in a macroscopically occupied single quantum state, i.e., a BEC, evolving on the five-level Hilbert space given by the $F=2$ rubidium hyperfine ground state (Fig.\,\ref{Rb_levels}). 
The atoms are first loaded at room temperature in a ultra-high-vacuum glass cell by means of a pulsed dispenser. Then, an atom chip equipped with a reflective golden layer is mounted in the science cell to create, together with a pair of external Helmholtz coils, a mirror magneto-optical trap (MOT) that laser cool and trap the atoms. The latter are optically pumped in the $|F=2, m_F=+2\rangle$ level (Fig.\,\ref{Rb_levels}) and transferred to a magnetic micro-trap, generated by micro-structured conductors hosted on the chip, with longitudinal and radial trap frequencies of $46\rm\,Hz$ and $950\rm\,Hz$ respectively. Quantum degeneracy is reached by forced evaporative cooling, ramping down the frequency of a radio frequency field supplied by a waveform generator connected to a U-wire hosted by the chip. The BEC produced so far has typically $10^5$ atoms, a critical temperature of $0.5~\mu{\rm K}$ and forms at a distance of $300~\mu{\rm m}$ from the chip surface. All the subsequent manipulations described in the paper are performed $0.7~\rm{ms}$ after releasing the atoms from the magnetic trap; in this way, the cloud expansion guarantees bias field homogeneity and the effect of atomic collisions can be neglected. Hence, in the limit of independent atoms, recording the population distribution in each sub-level directly yields the probability for a single atom to occupy the latter. Moreover, an homogeneous and constant magnetic field, set to $6.179 \rm\, G$, is applied to the atoms to energetically separate the five sub-levels and define the quantization axis of the system. The value of the magnetic field is chosen such that it is much larger than magnetic noise fluctuations and, at the same time, the current (used to produce it) is not high enough to cause significant heating of the coils. The RF-field that drives the evolution of the BEC in the $F=2$ manifold described in the paper, is realized by another waveform generator connected to a second U-wire integrated on the chip. Note that the characteristic frequency for the free Hamiltonian evolution is of the order of $4~\MHz$, thus much faster than the controlled dynamics. Finally, the atoms distribution across the $F=2$ manifold is detected following a Stern-Gerlach method. After $1\rm\,ms$ of expansion, an inhomogeneous magnetic field is applied along the quantization axis for $10\rm\,ms$. The atoms move in the field gradient and their different $m_F$ states spatially separate. After a time of $23\rm\,ms$ of expansion, a standard absorption imaging sequence is executed. Since the imaging method is destructive, a new condensate has to be created each time and the measurement procedure repeated after different preparation states.

\end{document}